\documentclass{aa}  
\usepackage{color}

\usepackage{graphicx}
\usepackage{txfonts}
\begin{document}

   \title{Black holes as tools for quantum computing by advanced extraterrestrial civilizations}

   \author{Gia Dvali\inst{1,2}
          \and
          Zaza N. Osmanov\inst{3,4}
          }

   \institute{Arnold Sommerfeld Center, 
	Ludwig Maximilians University, Theresienstra{\ss}e 37,
	 80333 Munich, Germany \
              \email{gdvali@mpp.mpg.de}
        \and
             Max Planck Institute for Physics, F\"ohringer Ring 6, 80805 Munich, Germany
        \and     
              School of Physics, Free University of Tbilisi, 0183, Tbilisi, Georgia \
              \email{z.osmanov@freeuni.edu.ge}
              \and
              E. Kharadze Georgian National Astrophysical Observatory, Abastumani, 0301, Georgia
             }

   \date{Received ; accepted }

 
  \abstract
  
   {We explain that black holes are the most efficient 
 capacitors of quantum information. It is thereby expected that all sufficiently advanced civilizations ultimately employ black holes in their quantum computers.  
}
   {We use the methods developed for the study of black hole physics and by applying the Drake formula we estimate the observational characteristics. }
   {The accompanying Hawking radiation is democratic 
in particle species. Due to this, the alien quantum computers will radiate 
in ordinary particles such as neutrinos and photons
within the range of potential sensitivity of our detectors. 
 }
   {This offers a new avenue for SETI,  including the 
civilizations  entirely composed of hidden particles species
interacting with our world  exclusively through gravity.}
   {}


   \maketitle
%

\section{Introduction} 

The search for extraterrestrial intelligence (SETI) is one of the longstanding problems of humanity. It is clear that any technological civilization most probably should have certain technosignatures that might be potentially detectable by our facilities. The first SETI experiments started in the last century in the framework of radio listening to stars \citep{tarter}. Nowadays, the most advanced technology is the five hundred meter aperture spherical radio telescope FAST, which is dedicated to observing the sky, and one of the missions is to search for radio messages from extraterrestrial civilizations \citep{FAST}. In $1960$ Freeman Dyson proposed a revolutionary idea of searching for megastructures visible in the infrared spectrum \citep{dyson}. He has proposed that if an advanced alien civilization is capable of constructing a spherical shell around their host star to utilise its total emitted energy, the megastructure, having a radius of the order of $1$AU, will be visible in the infrared spectral band. This idea has been revived thanks to the discovery of the Tabby's star \citep{tabby} (characterized by flux dips of the order of $20\%$, automatically excluding the possibility of a planet shading the star). This stimulated further research and a series of papers have been published considering the Fermi's paradox \citep{fermi} the planetary megastructures \citep{type1}, the stability of Dyson spheres (DS) \citep{wright}; extension to hot DSs \citep{paper3,paper5}; megastructures constructed around black holes \citep{DSBH}; white dwarfs \citep{DSWD} and pulsars \citep{paper1,paper2}.

Since the search for technosignatures is very complex and therefore sophisticated, the best strategy for SETI should be to test all possible channels. In the light of this paradigm, it is worth noting that, on average, the solar-type stars in the Milky Way are much older than the Sun. Therefore, potentially detected technosignatures will belong to advanced alien societies. 

It is natural to assume that the technological advancement of any 
civilization is directly correlated with the level of  its computational 
performance.  This is fully confirmed by our own history. 
  Here,  we wish to point out that there exist certain universal markers 
 of computational advancement  which  
can be used  as new signatures in SETI.

In particular, in the present paper we wish to argue that any highly advanced civilization 
is expected to employ black holes for the maximal efficiency 
of their quantum computations.
The optimization between the information storage capacity   
and the time scale of its processing suggests that a substantial 
fraction of the exploited black holes must be the source of highly energetic Hawking radiation. 
 
 This gives new ways for the SETI.  It also provides us with new 
 criteria for constraining the parameters of highly advanced civilizations.
 
 The logic flow of the paper is as follows. 
  We first introduce the current framework of fundamental physics 
  and justify why the most basic laws of quantum theory and gravity must be shared by all ETI.  This applies to civilizations that may 
  entirely be composed by yet undiscovered elementary particle species
and experiencing some unknown interactions.  
  
   We then argue that for all civilizations, regardless their 
  composition,  black holes are the
   most efficient storers of quantum information. 
   This does not exclude the parallel use of other hypothetical objects 
   which within their particle sector maximize the information storage capacity, 
 the so-called ``saturons" \citep{DvaliSaturon}.  Following the previous paper, we explain 
 that among all such objects the black holes are the most efficient and 
 universal capacitors of quantum information. 
 
   Our conclusions are based on recent advancements 
   in development of a microscopic theory of 
  a black hole \citep{DvaliGomezNP, DvaliGomezCRIT, DvaliSCR}
  as well as on understanding the universal mechanisms of enhanced 
 information capacity \citep{DvaliArea, DvaliNeural, DvaliBHBrain, DvaliMichelZell, 
  DvaliPanchenko, DvaliMBurden, DvaliMETA} and  
  using them for ``black hole inspired" quantum computing in prototype 
  systems  \citep{DvaliPanchenko}.

     Especially important is the knowledge that among objects saturating the universal quantum field theoretic bound on microstate entropy,  black holes 
    posses the maximal capacity \citep{DvaliSaturon}.

 Distilling the above knowledge,  we argue that black hole based computer technologies represent an universal attractor points in the advancement path of any sufficiently long-lived civilization.  
 
  Relying solely on the laws of physics that must be shared by all 
  ETI, we draw some very general conclusions about their 
  usage of black hole computing.  
    The optimization of information storage capacity and 
    the time of its retrieval
suggests that it is likely that ETI invest their 
energy in manufacturing a high multiplicity of microscopic black holes, as opposed to few large ones.  
Correspondingly, their computers are likely to 
 produce a Hawking radiation of 
high temperature and intensity. 

  Another source of radiation of comparable frequency is expected 
  to come from the manufacturing process of black holes. 
  As we explain, this process likely requires the high energy particle collisions.

 We give some estimates of what the above implies for SETI. 
 In particular, considering the IceCube instruments (the most efficient neutrino observatory), based on 
 their sensitivity and the observed VHE neutrino flux in the energy domain $40$ TeV, we conclude that the mentioned observatory can detect emission from the black holes of advanced civilizations if their numbers are in the following intervals: $2.5\times 10^3\leq N^{\nu}_{_{II}}\leq 4.2\times 10^4$, $2\times 10^6\leq N^{\nu}_{_{III}}\leq 1.4\times 10^8$.

 Finally, we discuss the possibility  that nature houses a large number of 
 hidden particle species interacting with us 
 via gravity. We point out that because of the known universal relation 
 between the number of species and the mass of the lightest black holes 
 \citep{DvaliN, DvaliRediN}, the existence of many species 
 makes black hole manufacturing more accessible for  ETI. 
 
 This applies equally to our civilization. 
 In this light, a possible creation of microscopic black holes 
 at our particle accelerators, shall reveal a powerful information
 about the computational capacities of ETI.

The paper is organized as follows: in sections 2 - 10, we examine basic 
concepts of quantum computing by using black holes, in section 11 we discuss the observational features of such black holes, in sec.12 we discuss the hidden sector of particles and in sec.13, we summarise our findings.

\section{The framework} 

\subsection{Quantum physics and elementary particles}  

Our current framework for describing nature at a fundamental level
is based on quantum physics. 
  This framework passes all the tests, both theoretical
 and experimental,  available to our civilization.  
    The experiments prove that  the quantum laws work to the shortest distances directly probed at the human made accelerators.  In particular, the scale  $\sim 10^{-17}$cm has been reached at the Large Hadron Collider (LHC)  at CERN.  This extraordinary resolution is achieved in particle collisions at around $\sim$TeV energies. 
        
    In addition, the predictions obtained by the theoretical extrapolations of the quantum framework to much higher
 energies, are fully consistent with all the existing data.  A classic example is provided  by Quantum Chromo-Dynamics (QCD), which is the theory of the nuclear forces. 
 QCD is an  asymptotically-free quantum field theory, self-consistently 
valid at arbitrarily short distances/high energies. 

 This success by no means implies that our knowledge of fundamental forces is complete.
Certainly, new interactions can (and hopefully will) be discovered by the experiments with higher energies and/or higher precisions.  Many such interactions have been hypothesized and studied.  These are motivated by various open questions, such as 
the origin of dark matter, early cosmology, unification of forces and 
so on.   For none of these extensions, there exists even a slight evidence for non applicability of the basic laws of quantum physics.  
 Even the string theory, which theoretically culminates the idea of unification of all known interactions, fully relies on the laws of quantum physics. 
  
 Of course, the particular realizations of the  quantum framework may encounter some computational difficulties in certain regimes.  
 This happens, for instance,  when the interactions among the elementary particles become strong. In such a case, the perturbation theory 
 in a given coupling constant may break down and one needs to 
 resort to non-perturbative techniques such as, for example, the lattice simulations.  
  None of these computational obstacles bare any indication of the invalidity of the basic concepts of the quantum theory.  Rather,  they only  signal the need for improvement of computational tools
without going outside of the quantum framework. 

  In the light of all the above, in our analysis we shall assume the validity of the laws of quantum physics for ETI. 
 \subsection{Gravity}

 The second powerful factor is gravity.  At the classical level, 
 the gravitational interaction is described by Einstein's General Relativity (GR). Again, this is a remarkably successful theory,
 tested by direct observations within an impressive range of distances,  
 $10^{-1} - 10^{28}$cm. In addition, various successes of the early cosmology suggest that the validity domain of GR extends to much shorter distances. 
 
    At the quantum level, the excitations of the gravitational field are gravitons, the massless particles of spin$\, =2$.  
  All the classical interactions of macroscopic sources can 
  be derived from the quantum theory in form the exchanges of virtual gravitons, 
  in the expansion in series of Newton's constant 
  $G$.     
    
  The unique property of graviton is that it is sourced 
  by the energy momentum tensor. 
   This feature is universal  and accounts for all contributions 
including the energy-momentum of the graviton itself. Due to this property, the gravitons self-gravitate. In other words, any graviton acts as a source for others.
         
 The quantum states of the isolated gravitons have never been tested directly in lab experiments. This is explained by an extraordinary 
 weakness of their coupling at laboratory scales. For example, an interaction strength 
of a graviton of $\sim$cm wavelength is suppressed
by $\sim 10^{-66}$ \citep{DvaliGomezNP}.  The strength of the classical gravitational field produced 
by the macroscopic sources is a collective effect of 
a large number of gravitons. This is why, it is much easier to detect the classical effects of gravity as opposed to 
its quantum manifestations.  
 
  The story however is scale-dependent.  In the quantized version of Einstein gravity, the interactions among the individual gravitons are weak at  distances larger than the Planck length, 
 \begin{equation} \label{LP}
 L_P \equiv \sqrt{\hbar G/c^3}\,, 
 \end{equation} 
 where $\hbar$ is the Planck's constant and $c$ is the speed of light.  The Planck length is approximately $10^{-33}$cm.  At such separation, the characteristic momentum-transfer among the individual gravitons is of order the Planck momenta, $\sim M_Pc$, where 
  \begin{equation} \label{LP}
 M_P  \equiv \frac{\hbar }{c L_P} \,, 
 \end{equation} 
 is the Planck mass.
    The quantum gravitational effects become strong at this scale. 
   Due to this, $L_P$ is commonly considered to be an ultimate cutoff of the Einstein theory of gravity, viewed as effective quantum field theory.

  However, in any extension of the pure Einstein gravity,
 the actual length-scale, $L_*$, at which quantum gravity becomes strong,  is longer than $L_P$. 
 In general,  if a theory includes $N_{\rm sp}$ distinct elementary particle species, we have \citep{DvaliN, DvaliRediN}, 
  \begin{equation}\label{Lbound}
  L_* = \sqrt{N_{\rm sp}} L_P\,.
 \end{equation}     
   The corresponding scale of the momentum-transfer,  $M_*c$, is respectively lowered to  
    \begin{equation}\label{Mbound}
  M_*c= \frac{M_Pc}{\sqrt{N_{\rm sp}}} \,.
 \end{equation}               
 This scale is usually referred to as the ``species scale". 
  
  The scale $L_*$ imposes an absolute lower bound on compactness.  In particular, the localization radius   
 of a quantum information bit is bounded by $L_*$
 \citep{DvaliGomezQI}. 
  
  The currently known elementary particle species include 
 the degrees of freedom of the Standard Model (quarks, leptons, Higgs and gauge bosons) and the graviton.  
  Counting all physical polarizations, this amounts to $N_{\rm sp} \sim  100$.  Correspondingly,  the theory-independent bound on the compactness cutoff is $L_* \gtrsim 10 L_P \sim 10^{-32}$cm.

\subsection{Universal criteria obeyed by ETI}  
  
   Putting all the knowledge together, we cannot assume 
that an advanced ETI necessarily shares the elementary particle composition with us.  In particular,  some ETIs may originate from entirely different  particle species, interacting with us only gravitationally.  
   
  In the light of the above, we shall make the following (maximally-conservative) assumption: \\
 
  {\it   We share with ETI the laws of quantum physics
 and gravity. }  \\
 
 One of the consequences of this is that the 
 physics of black holes is common, irrespectively by which ETIs they are manufactured.  This shall 
 play the crucial role in our idea of SETI.

\section{Optimization of information storage}  

 The two essential ingredients of any type 
  of computation are hardware and software. 
We shall not attempt to make any assumptions about the 
software  used by the advanced ETI. The power of 
programming and the sophistication of their algorithms
are most likely way beyond our imagination.     

However, since the laws of quantum physics and gravity are common, we can make certain conclusions about the 
limiting capacities of their hardware. 
 We shall parameterize these limits according to certain well-defined characteristics. The main features we will be interested in are
the efficiencies of the quantum information storage 
and of its retrieval.  These parameters are not payed 
much attention at the present stage of development
of the quantum computing by our civilization, due to more urgent 
technological issues.   However,  they must become crucial 
for any civilization that is able to approach the limiting
capacities. 

In order to understand what this limits imply, we go in two steps.  First, in the current  section we define what it means for a generic device to have an enhanced capacity of information (memory) storage and what is the underlying mechanism behind it.
  That is, we shall reduce the feature of the enhanced capacity of information storage to its bare essentials
 borrowing the  results of the series of papers
\citep{DvaliArea, DvaliNeural, DvaliBHBrain, DvaliMichelZell,  DvaliMBurden, DvaliMETA}. The reader can find a self-contained summary and discussion in \citep{DvaliMETA}.  

  Next,  we shall explain that among all possible  hypothetical 
 devices that saturate the information storage 
  capacity,  the black holes are the most efficient ones
  \citep{DvaliSaturon}. 
   
In general, the quantum information is stored in a quantum state of the system.  These states are usually labeled by the excitation levels 
of a set of $N$ elementary degrees of freedom. 
Their choice is a matter of convenience depending on the system parameters, available technology and the computational task. 
The choice may range from atomic spin states, for less advanced 
ETI, all the way to graviton excitations that can be used by highly advanced  civilizations.   
Regardless of their particular nature,  it is customary to describe them 
as quantum oscillators in the number representation. 
Following the terminology of 
\citep{DvaliArea, DvaliNeural, DvaliBHBrain, DvaliMichelZell,  DvaliMBurden, DvaliMETA},  we shall refer to these as the ``memory modes".

 The simplest mode is a qubit, a quantum degree of freedom that can exist 
in two basic states.  These states can be labelled as the occupation number
eigenstates $|0>$ and $|1>$ of the corresponding quantum oscillator with a number operator $\hat{n}_j$, and $j=1,2, ...,N$ the 
mode label.   

 Correspondingly, the set of $N$ qubits produces $n_{st} = 2^N$ basic states. These represent the tensor products of individual 
basis states  of $N$ distinct memory modes, $|n_1,n_2, ...,n_N> 
\equiv |n_1>\otimes |n_1>\otimes, 
..., \otimes |n_N>$, where $n_j = 0$ or $1$, for each 
$j$. 
In short, they count all possible sequences of $0$-s and $1$-s, such as, $|0,0,...,0>, 
~|1,0, ...,0>,~..., |1,1, ...,1>$.  We shall refer to these states as the ``basic 
memory patterns".  They form a Hilbert (sub)space, which we shall call 
the ``memory space". 

The system can exist in an arbitrarily entangled 
state of the memory modes.   A typical entangled  state would represent a linear superposition 
of order-one fraction of $n_{st}$ basis states.  However, there can exist 
the maximally entangled states of special types, such as the
generalized EPR state,  $\frac{1}{\sqrt{2}}(|0,0, ...,0>  + |1,1, ...,1>)$.

 As a measure of the information storage capacity of the system,  it is convenient to use the microstate entropy. 
As usual, we define it as the log from the dimensionality of the memory space, 
 \begin{equation} \label{entropy}  
   S = k_B\ln(n_{ts}) \,. 
 \end{equation}
We shall focus on two important characteristics of the  
efficiency of a quantum memory device:   
 The energy-efficiency and the compactness.

 The first one is determined by the energy gap of qubits, 
 $\epsilon$. This also determines  the total energy gap
 occupied by all $n_{st}$ memory states, which we denote by 
 $\Delta E$.   
 
 For example, if for each degree of freedom the states 
 $|0>$ and $|1>$ are 
gapped by $\epsilon$,  the gap spanned by the memory space  
will be $\Delta E = N\epsilon$. 

 Another important parameter is the radius of localization, $R$, of the quantum information storing device.  
 Ordinarily, the compactness of the system 
 increases the gap. For example,
consider a qubit realized by a
relativistic particle, say a photon,  localized within a box of size $R$. The two basis states can for example be chosen as the states with one and zero photons.  If the interactions of the photon 
inside the box are weak, the minimal energy gap  due to quantum uncertainty  is
 $\epsilon \sim c\frac{\hbar}{R}$.   
 Consequently, the localization of $N$ such qubits within the same box,
  will result in a basic energy cost  $\Delta E \sim N\frac{\hbar c}{R}$.  
  
  The first necessary requirement fulfilled by the device of the enhanced memory capacity is to minimize the 
gap of the memory qubits below the ``naive" quantum uncertainty,   
\begin{equation}  \label{C1}
  \epsilon  \ll  \frac{\hbar c}{R}\,,
\end{equation}    
  while maximizing their number $N$ (equivalently, $S$). 
  
The crucial point is  that the two are not independent:  for a system of given 
compactness $R$ there exists an absolute upper bound on $S$ imposed by unitarity \citep{DvaliSaturon}.
 We shall discuss this bound later. 
 
  At the same time, the energy gap of a quibt $\epsilon$, 
  sets the lower bound on the time-scale required for processing
  the information stored in that qubit, 
  \begin{equation} \label{tq}  
   t_q = \frac{\hbar}{\epsilon}  \,. 
 \end{equation}
 In particular, this is the minimal time required for extracting the information from the qubit. 
 For example,  (\ref{tq}) is 
 the minimal time required for detecting a photon of energy $\epsilon$. 
 
 Notice that $\epsilon$ stands for an effective energy gap in which  the contributions  due to interactions with the environment, with other qubits and other parts of the device, are already taken into account. 
 
 The expression (\ref{tq}) shows that there is a tradeoff. 
 The cheap storage of information requires a longer time 
 for its retrieval. This shall play an important role 
 in estimating the computational parameters of ETI.

 At the earliest stages of the development of quantum computers, 
 as it is currently the case with our civilization, the pressing issues related to the hardware are the problems of  decoherence. 
 The questions of the energy efficiency of qubits, posed above, are secondary and not payed much attention.     
This makes sense, since even for highly optimistic numbers of the memory qubits, the energy costs due to the minimal 
quantum uncertainty are insignificant. 

For example, for $N \sim 10^{10}$, the energy gap of the 
memory space for a modern-technology information-storing    
device of size $R \sim $cm, is only $\Delta E \sim 10^5$eV. This is less than the 
rest energy of a single electron. Obviously, this is negligible  
as compared to the rest energy of a typical supporting device that 
consist of the macroscopic number of atoms.   

However, the amount of quantum information processed by an 
advanced civilization must require uncomparably larger values of 
$N$.  
  Therefore, an advanced civilization that cares about the compactness and the energy efficiency of its computations must find the way of optimizing the information storage capacity of the system.  That is, it has to manufacture devices 
  that maximizes the microstate entropy per given size of the system.
   Basically, the goal is to maximize $S$ for given $R$.  
    
  Let us now explain that, for achieving the above goal,
the sufficiently advanced ETIs are expected to use black holes.

 \section{Why Black Holes?}    
 
 Black holes exhibit an extraordinary capacity of the information storage
 and processing. For various comparisons, it is illuminating to take the human brain as a reference point 
 \citep{DvaliBHBrain}. For example,   
 a black hole of a mass 
 of a human brain ($\sim $kg) has a million times larger memory capacity 
 while being in size only $\sim 10^{-25}$cm  and having  
 the operational time of memory read-out $t_q \sim 10^{-19}$sec.
  This is impressive, but are they the optimal tools? 
 
In order to classify black holes as the ultimate device 
used in quantum computing by advanced ETIs, the following set of 
questions must be answered:   
 
 \begin{itemize}

 \item How unique are black holes in their 
  ability of information storage?  
  
  \item What are the general mechanisms 
 behind this ability? 
 
 \item Can these mechanisms be implemented 
 in other devices for achieving 
  a comparable, or even a higher, efficiency of information storage and processing?
  
  \end{itemize}

Fortunately, these questions have already been largely addressed in
a research program consisting of three parallel 
strategies: 

\begin{itemize}
  \item Formulate a microscopic theory of a black hole
  \citep{DvaliGomezNP}.
 Demonstrate that the analogous mechanisms of information storage 
 and processing can be implemented in other quantum systems \citep{DvaliGomezCRIT, DvaliSCR, DvaliGOLD, DvaliPanchenko}.   
  
  \item Use the well stablished features of a black hole in order to deduce the properties of its memory modes, such as, the energy gaps 
  of qubits, their diversity and the evolution time scales.    
 Next,   
identify the mechanisms in generic quantum systems leading to 
the similar features. Implement such systems for quantum information processing \citep{DvaliPanchenko, DvaliNeural, DvaliBHBrain, DvaliArea, 
DvaliMBurden, DvaliMichelZell, DvaliMETA}
  
  \item Identify objects in generic quantum field theories that are equally (or more) efficient in information storage and processing as black holes.  Extract the universal
underlying physical reasons behind these features \citep{DvaliSaturon}.
  
\end{itemize}

   Remarkably, all three strategies have converged on 
   one and the same picture of the properties of the black hole memory modes (qubits) and 
  the mechanism behind their gaplessness and abundance. 
  
  In particular, it turns out that these features are generic for the objects saturating 
  a certain universal upper bounds on the microstate degeneracy imposed 
  by unitarity \citep{DvaliSaturon}.   The objets are called ``saturons". 
 Since, the bound is universal, it is shared 
  by all sectors of the universe, regardless their particle 
 composition. In particular, it is applicable even for yet undetected hypothetical particle species from the hidden sectors.   
  
   A long list of such objects has been identified in various consistent quantum field theories \citep{DvaliS1, DvaliS2, DvaliS3, DvaliS4, DvaliS5}.
  Interestingly, 
  they evidently exist in already discovered interactions,
 such as QCD, in form of color glass condensate of gluons
 \citep{GiaRaju}. 
  
   The universal mechanism of enhanced information storage capacity can be potentially implemented in laboratory settings, such as cold atoms.  This is a realistic prospect even for our civilization in not far future.

    The bottomline of this discussion is that an advanced civilization can certainly use non gravitational saturons for the efficient information storage. 
   
   Depending on the perspective, this may be regarded as good news or bad news. On one hand, it is exciting to be able to exploit  the black hole mechanism of information storage without the need of manufacturing the actual black holes, which for our civilization can be a long prospect.  Instead, we can  implement the  same mechanism in  ordinary systems such as cold atoms or the nuclear matter. 

 On the other hand, one may worry that this option makes the signatures for SETI less universal and more technology dependent.    
       
   However,  we can claim with certainty \citep{DvaliSaturon} 
   that among all possible saturons, the black holes 
 stand out as the most efficient information storers. 
 This gives us a confidence that a highly advanced ETI 
 must be using them as the information storing device. 
  
  This provides us with a prospect of search for advanced civilizations 
 even if they are composed out of entirely unknown particle species.  For example, ETI may consist of particles from a 
 hidden sector not sharing any constituents with the
 species that are known to us. 

 Nevertheless, 
 the black hole quantum computing device of such a hidden civilization  will provide signatures  detectable by our  observers.  This is because Einstein gravity is universal for all the particle species.  A direct consequence of this fact is the universality of the Hawking radiation.     
   Correspondingly, the quantum computers of dark ETI will 
  radiate democratically into the particles  of all the species, including the particles of the Standard Model which we can potentially detect.   
 
 \section{Closer Look at Black Holes} 
 
 \subsection{Generalities} 
  
  Black holes are the most compact objects of nature. Their effective size is given by the gravitational radius $R$. For a non-rotating and electrically-neutral black hole, this is set by
  the Schwarzschild radius,  $R= 2GM/c^2$. 
  
  In the classical description,  black holes are stable and eternal. 
 In quantum picture the story changes. 
 Due to quantum effects, the black holes evaporate by emitting 
   Hawking radiation.  Hawking's original calculation \citep{hawking}
  was performed in exact semi-classical limit. This 
implies taking the black hole mass to infinity and Newton's constant to zero, while keeping the black hole radius fixed, 
 $G =0,~M=\infty,~ R=$finite.  In this limit, the back reaction from the emitted quanta vanishes and the emission rate can be evaluated exactly.  The resulting spectrum of radiation is thermal, with an effective temperature given by  
  \citep{hawking},
\begin{equation}
\label{T} 
T = \frac{\hbar c^3}{8\pi k_{_B}MG} \,.
\end{equation}
This leads to the following Stephen-Boltzmann law 
\begin{equation}
\label{dmdt} 
c^2 \frac{dM}{dt} = - N_{\rm sp} 4\pi\sigma R_{_{H}}^2T^4 \,.
\end{equation}
Here  $N_{\rm sp}$ counts all particle species (all polarizations) 
lighter than $T$. That is,  all species are produced with an universal  thermal rate. 
The particles heavier that $T$ are Boltzmann suppressed and  can be ignored.   
 The democratic production of all particle species is one of the important 
 properties of the Hawking radiation. This feature originates from the universal nature of the gravitational interaction.  
 
Of course, in Hawking's computation the black hole has infinite mass and therefore it evaporates eternally.

  For a finite mass black hole, the story is 
different.  In order to estimate the decay time, one needs 
to make some additional assumptions.
The most common is to assume that the black hole evaporation is self-similar. 
That is, the relation between the mass and the temperature (\ref{T}) holds throughout the evaporation process. Under this assumption, 
the system can be easily integrated and one gets the following expression 
for the half-decay time,   
\begin{equation}
\label{tau1} 
\tau_{_{1/2}} = N_{\rm sp}^{-1} \frac{4480\pi G^2 M^3}{\hbar c^4} sec
\end{equation}
   For getting a better feeling about time scales, it is useful to rewrite this
   as an approximate expression,      
\begin{equation}
\label{tau2} 
\tau_{_{1/2}} \simeq 0.7 \times\frac{100}{N_{\rm sp}}\times\left(\frac{M}{10^9\; g}\right)^3 sec \,.
\end{equation}
 The factor $\sim 100$ in the numerator stands for a normalization 
 relative to the number of known particle species, while 
 $N_{\rm sp}$ refers to a total number of species available 
 at initial temperature of a black hole.

  However, it is important to keep in mind that for the finite mass black holes, the thermality is only an  approximation, since the back reaction is finite. 
  By the time the black hole looses half of its mass,  
   both expressions are expected to receive the substantial corrections due to quantum back reaction. 
   This is evident from various sources:
   the general consistency arguments \citep{DvaliT}, 
   the detailed analysis of prototype quantum systems 
  \citep{DvaliMETA}, as well as, 
   from the explicit microscopic theory of a black hole \citep{DvaliGomezNP}.
   It is independently evident that due to the back reaction from 
  the carried quantum information,  the black hole evaporation cannot stay self similar throughout the process, and certainly not beyond its half-decay \citep{DvaliMETA}.

This knowledge is no obstacle for our analysis.         
 For the conservative estimates,  the above expressions 
  can be used as good approximation till the black hole half-decay time.   This self-consistently 
 shows that the expression  (\ref{tau2}) gives a correct order of magnitude approximation for this time scale. 
 This is fully sufficient for our purposes. Our discussions will be limited by this time-scale.  
 
 Let us now discuss the information storage capacity of black holes.   
   It is well known that a black hole of radius $R$  has a microstate entropy equal to its surface area measured in units of 
   $G$  \citep{Bekenstein-Hawking}, 
\begin{equation}
\label{S} 
S = k_B\frac{\pi c^3R^2}{\hbar G} = k_B\frac{\pi R^2}{L_P^2} \,,
\end{equation}
where in the last part of equation we have expressed it through the  Planck length. 

Another important feature of a black hole is the information
retrieval time.  The huge entropy of a black hole shows that it 
can carry a large amount of information. During the evaporation, 
the black hole looses its mass. However, initially a very little information comes out. This is because at initial stages of decay, the spectrum 
of Hawking radiation is thermal  within a very good approximation.  
  The quantum information about the black hole state is encoded in 
  deviations from thermality, which are of order $1/S$ per one emission time 
  \citep{DvaliGomezNP, DvaliT}.  Initially, the effects of these corrections are  small and 
  it takes a very long time to resolve them.  

Page argued \citep{Page} that the start of the information retrieval 
from a black hole is given by its half decay time (\ref{tau2}).  
   As we shall discuss below, this feature has a physically transparent 
  explanation in terms of gaps of the black hole qubits \citep{DvaliGomezNP}. 
    It also was shown that the same feature is shared by 
  all objects of maximal microstate entropy \citep{DvaliSaturon}.

  Taking (\ref{S}) as a fact, one can deduce an useful knowledge 
  about the information-storing qubits without  making any 
  assumptions about their nature \citep{DvaliArea, DvaliBHBrain, DvaliNeural,  DvaliMETA, 
  DvaliGOLD}.  
  In particular, it is clear that the number of qubits is 
  $N \sim S/k_B$.  From here it follows that the energy gap of black hole qubits is around, 
   \begin{equation}
\label{gap} 
 \epsilon = \frac{k_B}{S}\frac{\hbar c}{R}
 = \frac{\hbar^2 G}{\pi c^2 R^3} \,.
\end{equation}
Notice that for $N_{\rm sp} \sim 1$,  
the corresponding information storage
 time-scale (\ref{tq}), 
    \begin{equation}
\label{time} 
 t_q = \frac{\hbar}{\epsilon} = \frac{S}{k_B} \frac{R}{c}
 = \frac{\pi c^2 R^3}{\hbar G} \,,
\end{equation} 
 is of the same order as the half-decay time (\ref{tau2}). 
 Thus, we have, 
     \begin{equation} \label{equality} 
 t_q \sim \tau_{1/2} \,.
\end{equation} 
 This nicely reproduces the Page's estimate. It also shows 
 that this estimate is a straightforward consequence 
 of the energy gaps of the black hole qubits. 
 
 Of course, for $N_{\rm sp} \gg 1$, the decay time is shortened and so is the time-scale of information processing,
     \begin{equation}
\label{timeN} 
 t_q = \frac{1}{N_{\rm sp}} \frac{S}{k_B} \frac{R}{c}
 = \frac{1}{N_{\rm sp}} \frac{\pi c^2 R^3}{\hbar G} \,,
\end{equation}

 \subsection{Cross-check  from a microscopic theory} 
   
   The expressions (\ref{T}), (\ref{S}), (\ref{gap}), (\ref{time}),  (\ref{timeN}) have been independently derived within a microscopic theory of a black hole, 
   the so-called ``quantum $N$-portrait" \citep{DvaliGomezNP}.
   This theory tells us that at the fundamental level 
   a black hole is described as 
 a bound-state of $N \sim S$ gravitons of wavelengths $\sim R$. 
 This collection of gravitons can be regarded as a coherent state or a condensate.  
  
 The black hole qubits originate from the collective 
 Bogoliubov-type excitations of the graviton condensate.
  As shown in \citep{DvaliGomezCRIT}, and further studied 
  in series of papers \citep{DvaliSCR, DvaliGOLD, DvaliPanchenko, 
  DvaliNeural, DvaliArea, DvaliBHBrain, DvaliMBurden, DvaliMichelZell, DvaliMETA},  a similar emergence 
  of gapless qubits is characteristic to systems of attractive 
 Bose-Einstein condensates at quantum critical points.  
       
 For a black hole, the qubits represent the excitations 
 of graviton with high angular momentum, which become gapless due to attractive interaction with the condensate.
  These excitations can be described in several equivalent 
  languages. For example, they can be viewed as Goldstone modes of the symmetries 
  that are spontaneously broken by a black hole. 
  Their emergence can also be understood in terms of 
  the mechanism of ``assisted gaplessness" \cite{DvaliArea}. The essence of this phenomenon is that the negative 
  energy of the attractive interaction compensates the positive 
 kinetic energy of the would-be free mode.  
  
   The independent derivation of the relations (\ref{T}), (\ref{S}), (\ref{gap}), (\ref{time}), (\ref{timeN}) from the microscopic theory provides an 
   important cross-check for the validity of these equations.

\section{Competing Devices: Saturons}

Let us now ask:  How do we know that there exist no  
information-storing devices competing with black holes?
For a long time, it was implicitly assumed that the area form of the entropy  (\ref{S}) was exclusively the property of black holes. 
However, this understanding has changed recently \citep{DvaliSaturon}. 
It has been shown that 
the area-law expression is common for all field theoretic objects
that have maximal entropy permitted by unitarity.  
 Namely, for a localized object of radius 
 $R$,  there exists the following universal upper bound on the microstate entropy \citep{DvaliSaturon}, 
 \begin{equation}
\label{Ssat} 
S = k_B\frac{\pi c^3 R^2}{\hbar \tilde{G}} \,,
\end{equation}
where $\tilde{G}$ is the coupling of the Nambu-Goldstone mode of spontaneously broken Poincare symmetry.  This mode is universally present 
for an arbitrary localized object, since any such object breaks the Poincare symmetry spontaneously.  Due to this, 
$\tilde{G}$ represents a well-defined 
parameter for an arbitrary macroscopic object. 
 The objects saturating the above bound are referred to as ``saturons"
 \citep{DvaliSaturon}. 
 
 Various explicit examples of such objects have been constructed in non-gravitational quantum field theories, including some candidates 
of cold atomic systems\citep{DvaliS1, DvaliS2, DvaliS3, DvaliS4, DvaliS5}. 

 An important finding of these studies is that all saturons 
exhibit the same information processing properties as black holes, with the role of $G$ taken up by $\tilde{G}$.  
Namely, all saturons decay by emission of Hawking-like radiation
with temperature given by (\ref{T}), 
\begin{equation}
\label{Tsat} 
T = \frac{\hbar c }{4\pi k_{_B}R} \,.
\end{equation}
The energy gap of saturon quibts is,
  \begin{equation}
\label{gapsat} 
 \epsilon = \frac{k_B}{S}\frac{\hbar c}{R}
 = \frac{\hbar^2 \tilde{G}}{\pi c^2 R^3} \,,
\end{equation}
and the corresponding information recovery time is, 
   \begin{equation}
\label{timesat} 
 t_q = \frac{\hbar}{\epsilon} = \frac{S}{k_B} \frac{R}{c}
 = \frac{\pi c^2 R^3}{\hbar \tilde{G}} \,.
\end{equation} 
 Of course, unlike black holes, the generic saturons radiate only in particle species to which their constituents interact.
 
It is certainly conceivable that some 
civilizations use non-gravitational saturon devices, as opposed to 
black holes,  for their quantum computations. 
In fact, even our civilization may not be too far from implementing the saturated systems for such a purpose. 

First, soon after the formulation of black hole $N$-portrait, the existence of similar gaplessness mechanisms were established in the prototype $N$-boson models \citep{DvaliGomezCRIT, DvaliSCR, DvaliGOLD} . 

Further, it has been proposed 
\citep{DvaliPanchenko, DvaliNeural, DvaliBHBrain, DvaliMichelZell} that 
critical Bose-Einstein condensates, 
can be used for implementing this black hole mechanism 
 for controlled information processing. 

 More recently, 
it has been argued \citep{GiaRaju} that saturated states have already been created in laboratory experiments in form of the color glass condensate of gluons \citep{CGC}.  It is not necessarily too far-fetched 
to expect the use of such states for quantum information processing on 
a reasonable time-scale of the advancement of our civilization.

\section{Can ETI use other saturons instead of  black holes?} 

In the above light, one may wonder whether the existence of saturons as of alternative information storing  devices may impair our idea of SETI.  The reason is that the radiation 
signature from such quantum computers will not be universal and will rather depend on a particle composition of ETI.  
 Fortunately, this is not the case.

The point is that  among all saturated objects of any given size 
$R$, the black holes are unbeatable 
 in their information storage capacity \citep{DvaliSaturon}.  This is due to the fact that the spontaneous breaking of Poincare symmetry is maximized by a black hole.  Correspondingly, the coupling of 
 the Poincare Goldstone in case of a black hole is the weakest.   
 That is, for any other object 
 of the same size, we have $\tilde{G} \geqslant G$. 
 In other words,  the non-black-hole saturons, irrespective of their 
 composition, must have $\tilde{G} > G$.   Any decrease 
 of  $\tilde{G}$ below $G$, will collapse the saturon into a black hole. 
 
 The details of the proof can be found in the 
 original article \citep{DvaliSaturon}. However, due to an interdisciplinary nature of the present paper, 
 for the sake of  the reader's comfort,  we shall make the presentation maximally self-contained.  For this purpose, we provide the following 
 physical explanation. 
 
  Imagine we would like to produce a saturated device 
  of size $\sim R$ in form of a self-sustained bound state 
  of  $N$ quanta of the wavelengths $\sim R$.     
    Let us assume that the dominant
  interaction responsible for binding the system has an interaction 
  strength set by a dimensionless coupling 
  $\alpha$.  That is,  the attractive potential energy among the pair 
  of particles is $V = - \alpha \hbar c/R$.   
 
  Of course, at the same time these particles
  interact gravitationally, with the strength $\alpha_{\rm gr} 
  = \frac{\hbar G}{c^3R^2} = \frac{L_P^2}{cR^2}$.  
   The physical meaning of this parameter is easy to understand
 if we recall that the energies of quanta are $\sim \hbar c/R$
 (see, e.g.,  \citep{DvaliGomezNP}).    
  However, we have assumed that gravity is subdominant
  with respect to the binding force of coupling $\alpha$.

   It is clear that in order 
  to form a bound state, the kinetic energy of each quantum
 $\frac{c\hbar}{R}$ must be 
  balanced by the potential energy of attraction from the rest,
 $V = \alpha N c \hbar/R$.    
 This gives an equilibrium condition 
  $\alpha N \sim 1$.

  Now, the bound state breaks Poincare symmetry spontaneously. 
  It breaks both the translations and the Lorentz boosts. 
  The order parameter for breaking of Poincare symmetry  is $\hbar N/\pi c^3R^2$, and the Goldstone coupling is $\tilde{G} = \pi c^3\frac{R^2}{\hbar N}$. 
  Then, according to the unitarity bound (\ref{Ssat}),
  the maximal possible entropy of such a bound state  is $
  S = k_BN$.  Let us assume that 
  this entropy is attained.  Can it exceed the entropy of the same size black hole (\ref{S})?

 This requires making $\tilde{G} = \pi c^3 \frac{R^2}{\hbar N}$
  as small as $G$. But this implies that the gravitational coupling 
  $\alpha_{\rm gr}$  must become equal to the one of non-gravitational 
  binding force $\alpha \sim 1/N$. 
 Thus, we are pushed to equality $\alpha_{\rm gr} = \alpha$. 
   Remembering that the rest energy of the object is 
   $\sim N c\hbar/R$, it is easy to evaluate its gravitational radius, 
   which comes out equal to $\sim R$. 
      
  It is clear that
  at this point the object is within its gravitational radius and 
  thus is a black hole. That is, any attempt of pushing the entropy 
  of the saturon of size $R$ beyond the entropy of the same size 
  black hole,  collapses the saturon into a black hole.

  Correspondingly, if the civilization is sufficiently advanced,
 for information processing, it will ultimately resort to black holes rather than to  other saturons.  
  
  This gives an exciting prospect of searches for advanced civilizations 
  based on a detection of Hawking radiation from their quantum computers. 
   This applies equally to the civilizations belonging to hidden 
   particle species, interacting with us exclusively via gravity.  
  Regardless, due to its universal nature,  the Hawking radiation 
  from their quantum computers will inevitably contain 
  our particle species. 
  
 \section{Some remarks on black hole manufacturing technology} 
 
 An universal prescription for a creation of a black hole of mass 
 $M$ is that the energy $E = Mc^2$ must be localized within 
 the sphere of the corresponding Schwarzschild radius 
 $R = 2MG/c^2$.  
 
 Of course, we cannot make a precise prediction of the methods 
 used by an advanced civilization for achieving this goal. 
 However, it is possible to outline some expectations 
 depending on the masses of black holes manufactured by ETI.  

   In particular,  we wish to separate the two regimes of black hole formation, to which we shall refer as ``soft" and ``hard" respectively. 
   
    If the mass of a black hole is sufficiently high, it can be manufactured softly, by putting together a low density of many 
non-relativistic particles and letting them 
to collapse.  This is essentially how an ordinary stellar collapse 
works. 

 However, for creation of microscopic black holes this method 
 may not be used.  Instead, one needs to accelerate particles 
 to high energies and collide them with small impact parameters.   Such collisions are usually referred as ``hard" in the sense that the 
momentum-transfer per colliding particle exceeds their rest masses. 
 
 Let us explain the difference on a concrete example.   
 In order to produce a solar mass black hole ($R \sim$km), it is sufficient to place of order $10^{57}$ non-relativistic neutrons within the proximity of few km.  Such a system will collapse into a black hole without the need of any hard collision among the neutrons. 

      At the same time, if we wish to produce a black hole of
the size of a neutron $R \sim 10^{-14}$cm, this method will not work.  
 The mass of such a black hole is approximately $10^{38}$GeV, which is $10^{38}$  times the rest mass of a neutron.  We thus need to localize  about $10^{38}$ neutrons within a Compton wavelength of a single neutron.  
 This cannot be done softly due to the resistance from the nuclear forces.  For example, a chunk of a nuclear matter 
 consisting of $10^{38}$ neutrons will not collapse easily into a black hole, due to 
 Fermi pressure and other factors.  
 
 Instead, in order to achieve the goal, the  
 neutrons have to be accelerated to ultra-relativistic 
 energies and collided with a very high momentum transfer. 

For example, in an extreme case we can accelerate 
two neutrons to a center of mass energy  of $\sim 10^{38}$GeV and 
collide them with an impact parameter 
$\sim 10^{-14}$ cm. 
 Of course, one can instead collide more neutrons each with less 
 energy. However,  for each choice the collision must be hard since the neutrons must be relativistic and exchange momenta exceeding (or comparable to)  their rest masses.

 This example shows the general tendency: 
 manufacturing the microscopic black holes, requires 
 hard collisions among high energy particles.  
 
 Of course, one can imagine an exotic situation when an 
 advance civilization has at its disposal a new type of 
 a heavy elementary particle, interacting
 purely via gravity. 
 For example, we can imagine a stable 
scalar particle of mass $\sim M_P$.  Notice, such a particle is essentially 
a quantum black hole, since  its Compton wavelength 
as well as its gravitational radius are both of order $L_P$
\citep{DvaliGomezNP}. 
Using this elementary building block,  the black holes of arbitrary masses can be produced  softly. 
 Indeed, an arbitrary number of non-relativistic 
  particles with proper initial velocities  can collapse into a black hole. Of course, upon crossing the Schwarzschild radius, the system always becomes relativistic. But the process is soft in the sense  that such particles need not exchange momenta exceeding their rest masses.  
 
 Putting such highly exotic possibilities aside, 
  it is reasonable to expect that for manufacturing
  small black holes, the advanced ETI use the method of 
  colliding highly energetic elementary particles.  
   
   Such collisions are expected to be accompanied by 
  the radiation of quanta of the 
  typical energy given by the inverse impact parameter. 
  Since the latter is of order the black hole radius,
  the expected energy of the radiated quanta is $\sim \hbar c/R$. This is comparable to the temperature of the Hawking radiation
  from a black hole that is the outcome of the collision. 
  The difference of course is that the collision radiation is neither 
  thermal nor democratic in particle species.    
    
     Due to this, its detectability, unlike the Hawking radiation, 
     will depend on the composition of particle species used 
     in collisions for manufacturing black holes.  
  
    In conclusion, the following general message emerges. 
  The ETI can also be searched through the primary 
  radiation emitted during the process of their manufacturing of 
  black holes. The characteristic energy is set by 
  the inverse impact parameter and is comparable to the energy 
  of the Hawking quanta emitted by a black hole manufactured in the 
  process.

    \section{Remarks on optimisation of information processing} 
 
  Another remark we would like to make concerns  the optimisation 
  of information processing by ETI.  Of course,
  we are not going to speculate about the details. Our point will be limited 
  by aspects that are imposed by laws of 
  physics which all ETI have to respect.  These are the laws of 
  quantum field theory and gravity. 
    These laws imply the relation (\ref{timeN})
    between the information extraction time 
    from a black hole and its radius/temperature and entropy.  
  The analogous relation holds for all saturons 
  (\ref{timesat}) \citep{DvaliSaturon}.
 
  This universal relation shows that for a single black hole 
 the information extraction time $t_q$ is fixed in terms of its 
 radius $R$ (or temperature, $T\sim 1/R$)  and $N_{\rm sp}$.
 Of course, from the same relation, the entropy of a black hole is also uniquely
 fixed in terns of $t_q$ and $N_{\rm sp}$.  
 Notice that the total number of species $N_{\rm sp}$ at a given scale 
 is the constant of nature, which ETI cannot change.
  The optimization problem must thereby be solved for the 
 fixed $N_{\rm sp}$.   
      
   This allows us to draw some general conclusions  about 
  a  likely arrangement of information storing devices by ETI.
 Namely,  for a given $t_q$, the only way of increasing the amount of stored 
 information, is by multiplying the number of black holes. 
   Increasing the entropies of individual black holes is not possible, 
   as this will  increase $t_q$. 
  
   As it is clear from (\ref{timeN}), for a system of $n$ black holes, each 
   of mass $M$,
    the information retrieval time $t_q \propto M^3/M_P^2$ is independent 
    of $n$.  One can thus arbitrarily increase the entropy of the system 
    by increasing $n$ without altering $t_q$. The total entropy grows as $n$ while  $t_q$ stays the same.     
   
    In contrast, taking a single black hole of the mass 
    $M'= nM$, increases the entropy of the system 
   by extra power of  $n$ relative to the total entropy of $n$ separated black holes. 
   However, simultaneously the information retrieval time 
   grows as the cubic power of $n$, 
   \begin{equation} \label{tqn}
   t_q \, \rightarrow \, t_q'\, = \, n^3\, t_q \,.  
   \end{equation} 
    Obviously, this compromises the information processing efficiency of the system.     
    Thus, the optimal way of increasing the information capacity 
    of the system for a fixed $t_q$ is to increase the number of black holes
    instead of investing the same total energy into a creation of bigger ones.

    We thus arrive to a very general conclusion:  \\
    
   {\it  The quantum computing 
    devices of advanced ETI are more likely to consist of multiple 
    small black holes rather than  a fewer big ones. }\\
    
      This implies that from black hole ``farms" of ETI, we are more 
    likely to expect the Hawking radiation of higher temperature and intensity.

\section{Time scales and messengers}

 The relation (\ref{timeN})  between the information processing time $t_q$ and 
 the black hole radius/temperature allows us to give some estimates 
 about the energies and the nature of the emitted Hawking radiation. 
  We must however note that such estimates change 
  if we move beyond Einstein gravity. 
   This departure is scale-dependent. 
  As already discussed, in any theory 
  with number of particle species $N_{\rm sp}$, 
  the gravitational cutoff is lowered to the species scale 
  $L_*$ given by the equation (\ref{Lbound}) \citep{DvaliN, DvaliRediN}.
    
  This length gives an absolute lower bound on the size 
 of a black hole tractable within Einstein gravity. 
   By taking into account the number of all currently known elementary 
  particle species ($N_{\rm sp} \sim 100$), the minimal size 
 of a black hole is still very short, 
 $L_* \sim 10^{-32}$cm.   
  
   The corresponding half-life of an Einsteinian black hole is 
  only slightly longer than the Planck time, $t_q \sim 10^{-43}$ sec.  
    This time sets an absolute lower bound on the information processing 
    speed offered by any ETI. 
    
    Here, for definiteness,  we shall assume that    
    $L_* \sim 10^{-32}$cm marks the validity domain of Einsteinian gravity.    
      A discussion of the cases with much larger number $N_{\rm sp}$ 
  will be postponed  towards the end of the paper. 
  However, here we wish to note that increasing 
  $N_{\rm sp}$ only helps in more efficient manufacturing 
  of ``artificial"  black holes.   This is due to the fact that 
  the threshold energy of black hole formation in particle collisions 
  is lowered according to (\ref{Mbound}) \citep{DvaliRediN}. 
  The size of the smallest black hole  (\ref{Mbound}) is correspondingly 
  much larger than in Einstein gravity with a small number of species.

  Correspondingly, existence of 
  large number of species makes the black hole quantum computing more 
  likely accessible for less advanced civilizations.
    In particular, for the limiting case of $N_{\rm sp} \sim 
    10^{32}$, manufacturing of smallest black holes will become (or 
perhaps already is) possible by our civilization. 
    
     Assuming for the time being that we are within the validity 
     of  Einstein gravity for the scales of interest, we can estimate 
    the parameters of the expected Hawking radiation for 
    certain reasonable time-scales.     
 
   For example, demanding that $t_q$ is not much longer than 
   a few seconds, we arrive to the conclusion that the energy of the 
   Hawking quanta is above $\sim $TeV energy. 
     With such a high temperature all the known particle
  species will be radiated by ETI black hole computers. 
  
   Of course, how efficiently the radiation 
   can reach us depends on number of factors. These factors include 
   the presence of a medium that can absorb or shield the radiation.  
   
   In general, since the neutrinos have the deepest penetration length, they  
   appear to be the most robust messengers. 
   In addition, the mediation by high energy gamma quanta
  and other species is certainly a possibility. 
  
    Of course, the high energy radiation coming from 
the distant ETI, will be subjected to the same universal cutoffs as the 
high energy cosmic rays produced by the natural sources.
 The example is given by the  Greisen-Zatsepin-Kuzmin (GZK)  cutoff 
 \citep{GZK1, GZK2}
 due to scattering 
 of high energy protons at the cosmic microwave background. 
 
 On top of the known natural factors, there can be unknown artificial 
 shielding effects. For instance, ETI may compose absorbing 
 shields around their black hole computers. 
  The motivation may vary from a simple defence against the 
  life-threatening high energy radiation, as it is the case for humans, 
  all the way to sophisticated  energy-recycling purposes. 
   These are the factors about which it is hard to speculate.    
   However, in their light the case for neutrino, as of a most promising 
  known messenger, is strengthened.

\section{Some general estimates}

  All the parameters required for our estimates - such 
   as the time scale of processing, the temperature of accompanying 
  radiation, the black hole mass and the entropy - can be expressed through one another.  
    The only extra quantity is $N_{\rm sp}$, which shortens  the processing time.  At the moment we assume that this quantity is 
    not much different from the number of known species. 
    Then, we are left with a single input parameter which we 
    shall take to be the information processing time $t_q$ (\ref{time}).
     As we already discussed (\ref{equality}), this time is approximately equal to a half decay time of a black hole $\tau_{1/2}$, 
  which shall be used below. 
           
We assume that the typical timescale which is required for computation is of the order of $1$ sec. Then, from Eq. {\ref{tau2}} one can show that the black hole mass will be of the order of $1.2\times 10^9$ g. In this context, an important issue a civilization has to address is energy, which is required for constructing an extremely compact object, $Mc^2\simeq 10^{30}$ erg.

In order to understand the capabilities of extraterrestrial civilizations, one should examine them in the light of the classification introduced by \cite{kardashev}. According to this approach, alien societies should be distinguished by their technological level. Type-I civilizations, in particular, use the entire power attained on a planet. In the case of the Earth, the corresponding value is of the order of $P_{_{I}}\simeq \frac{L_{\odot}}{4}\times\left(\frac{R_E}{R_{AU}}\right)^2\simeq 1.7\times 10^{24}$ erg s$^{-1}$, whereas the current power consumption is about $1.5\times 10^{20}$ erg s$^{-1}$, classifying our civilization as level $0.7$ \citep{sagan}. Type-II civilization utilizes the whole stellar power, $P_{_{II}}\simeq L_{\odot}$, and Type-III is an alien society, that consumes the total power of a host galaxy, $P_{_{III}}\simeq 10^{11}\times L_{\odot}$. Here, $L_{\odot}\simeq 3.8\times 10^{33}$ erg s$^{-1}$ is the Solar luminosity, $R_{AU}\simeq 1.5\times 10^{13}$ cm is the distance from the Earth to the Sun (astronomical unit - AU) and $R_E\simeq 6.4\times 10^{8}$ cm denotes the radius of the Earth. From these estimates it is obvious that even Type-I civilization can build a black hole with the aforementioned mass: it only needs to accumulate the energy incident on the planet during $\sim 7$ days.
 
In the search for extraterrestrial intelligence, the major task is to find specific fingerprints of their activity. Using black holes will inevitably lead to such observational features which, on the one hand, might be detected and, on the other hand, distinguished from typical astrophysical events. In particular,
we use the knowledge that black holes emit particles with energy in the range $(E, E+dE)$ with the rate \citep{hawking}
\begin{equation}
\label{rate} 
R(E,M)\equiv \frac{d^2N}{dtdE} = \frac{1}{2\pi\hbar}\times\frac{\Gamma_s(E,M)}{e^{\frac{8\pi GME}{\hbar c^3}}-(-1)^{2s}},
\end{equation}
where $\Gamma_s(E,M) = E^2\sigma_s(E,M)/(\pi\hbar^2 c^2)$, $s$ denotes the spin of a particle, $\sigma_s = \alpha_sG^2M^2/c^4$, $\alpha_{1/2} = 56.7$, $\alpha_{1} = 20.4$ and $\alpha_{2} = 2.33$ \citep{gamma}.  As discussed, the flux of particles might be screened out but, we assume that neutrinos escape. Therefore, the major observational feature might be the energetic flare of neutrinos. Using Eq. (\ref{rate}), one can easily demonstrate that the spectral power, $ER$, of neutrinos, ($s = 1/2$), has a peak at energies
\begin{equation}
\label{en} 
E^{\nu}_m\simeq\frac{3.13 \hbar c^3}{8\pi GM}\simeq 48\times\left(\frac{100}{N_{sp}}\times\frac{1\; sec}{\tau_{_{1/2}}}\right)^{1/3} TeV,
\end{equation}
which can be detected by the IceCube detector \citep{icecube}. For the neutrino emission luminosity at $E_m$ we obtain
$$L^{\nu}_m\simeq 3\times (E^{\nu}_m)^2R(E^{\nu}_m,M)\simeq$$
\begin{equation}
\label{power} 
\simeq 1.2\times 10^{28}\times\left(\frac{120}{N_{sp}}\times\frac{1\; sec}{\tau_{_{1/2}}}\right)^{2/3}\; erg/s.
\end{equation}
where the multiplier, $3$, comes from the three flavors of neutrinos. As it is evident, even a single event is characterized by the very high luminosity of the neutrino flare. In general, it is obvious that a civilization might perform many calculations per second. On the other hand, any civilization is limited by the level of technology it possesses. The maximum number of quantum processings per second, $n_x$ (where $x = (I, II, III) $ denotes the level of technology), is severely constrained by a civilization's maximum power, $n_x\simeq P_x/\langle L \rangle$, where $\langle L \rangle\simeq Mc^2/2\tau_{_{1/2}}$ is the average luminosity of black hole's emission:
\begin{equation}
\label{n1} 
n_{_I} = 3.5\times 10^{-6}\times\left(\frac{100}{N_{sp}}\right)^{1/3}\times\left(\frac{\tau_{_{1/2}} }{1\; sec}\right)^{2/3}\; sec^{-1},
\end{equation}
\begin{equation}
\label{n2} 
n_{_{II}} = 7.7\times 10^3\times\left(\frac{100}{N_{sp}}\right)^{1/3}\times\left(\frac{\tau_{_{1/2}} }{1\; sec}\right)^{2/3}\; sec^{-1},
\end{equation}
\begin{equation}
\label{n3} 
n_{_{III}} = 7.7\times 10^{14}\times\left(\frac{100}{N_{sp}}\right)^{1/3}\times\left(\frac{\tau_{_{1/2}} }{1\; sec}\right)^{2/3}\; sec^{-1}.
\end{equation}

Since the value of $n_{_I}$ is very small, the best way is to capture a single event of duration $\tau_{_{1/2}}$. For Type-I alien societies the luminosity will be defined by Eq. (\ref{power}), but for Type-II,III civilizations the luminosity might be much higher, reaching the solar luminosity for Type-II and the galactic luminosity for Type-III societies.

This is not the end of the story. In 1961, Frank Drake formulated an equation used to estimate the number of communicative civilizations \citep{drake}
\begin{equation}
\label{drake} 
N = R_{\star}\times f_p\times n_e\times f_l\times f_i\times f_t\times \mathcal{L},
\end{equation}
where $R_{\star}$ denotes the average rate of star formation, $f_p$ denotes the fraction of stars with planets, and $n_e$ denotes the average number of potentially habitable planets per star, $f_l$ is the fraction of planets, that can potentially support life and that actually develop it, $f_i$ denotes the fraction of planets with life that develop intelligent life, $f_t$ is the fraction of technological communicative civilizations, and $\mathcal{L}$ is the average length of a time-scale for which alien advanced civilizations release detectable signals into space. From the study of the Milky Way (MW), it is well known that the modern value of the star formation rate is in the interval $(1.5-3)$ stars per year \citep{rate}. Based on the Kepler space mission, it has been revealed that there should be around $40$ billion Earth-sized planets in the habitable zones of solar-type stars and red dwarfs \citep{fpne}. This automatically implies that a value of $f_p\times n_e$ should be $0.4$. The average value of $R_{_{Astro}} = R_{\star}\times f_p\times n_e\simeq 0.9$, is more or less well defined, but about the rest, almost nothing is known and we can only apply very speculative approaches. Based on the observation that life on Earth appeared soon after the conditions became favorable, one can speculate that the value of $f_l$ is close to $1$. In his statistical approach to the Drake equation, \cite{maccone} uses a moderate value of $1/3$ and a value of $0.01$ for $f_i \times f_t$. In the scientific literature, a range $(10^{-3}-1)$ \citep{prantzos} is considered for the quantity $f_{_{Biotech}} = f_l\times f_i\times f_t$. 

For estimating the value of $\mathcal{L}$, we assume that it should not be less than the timescale on which we need to reach the corresponding level of technology. Following an approach by \cite{dyson}, if one accepts that $1\%$ of the annual growth rate of industry is maintained, then by taking the current power consumpton, $P_0\simeq 1.5\times 10^{20}$ erg/s, into account, one can straightforwardly show that to reach Type-I and Type-II civilizations, one requires approximately $t_{I} = 1000$ yrs and $t_{II} = 3000$yrs respectively. Calculation of $t_{III}$ should be performed in a different way. By definition, a galactic civilization is the one which colonized the whole galaxy, therefore, the corresponding value should be of the order of $D_{MW}/\upsilon$, where $D_{MW}\simeq 26.8$ kpc is its average diameter of the MW galaxy and $\upsilon$ is a velocity by which a civilization covers the mentioned distance. By taking a very moderate value of it, $0.01 c$, one obtains $t_{III}\simeq 10^8$ yrs. It is clear that a theoretical minimum value of $t_{III}$ imposed by the speed of light is of the order of $10^6$ yrs.

By assuming that the minimum values of $\mathcal{L}_x$ are of the order of $t_x$, then the number of civilizations, $N = R_{_{Astro}}\times f_{_{Biotech}}\times\mathcal{L}$ can vary in the interval $(3,\;1.2\times 10^3)$ (Type-I) and $(6,\;3.6\times 10^3)$ (Type-II). The aforementioned values might be even higher, if  $\mathcal{L}_x$ exceed $t_x$. One should also emphasise that since nearly $75\%$ of stars with habitable planets are by $\sim 1$ Gy older than the Sun \citep{age}, the previously mentioned assumption is quite natural \citep{sagan2,cameron, shkl}, where the number of civilizations has been estimated to be of the order of $10^6$.

For the uniform distribution of civilizations in the galactic plane, one can find that the order of magnitude of a minimum distance between alien societies is given by $d_m\simeq\sqrt{A_{MW}/N}$, where $A_{MW}\simeq\pi D_{MW}^2/4$ is the area of the MW galaxy. Then, for the neutrino flux, $F^{\nu}\simeq L^{\nu}_m/(4\pi d_m^2)$, multiplied by $\pi^2\;n_{_{II}}$ for Type-II civilizations (multiplier $\pi^2$ comes from the summation over all civilizations in the galactic plane), one obtains
$$F^{\nu}_{_{I}}\simeq 10^{-13}\times\left(\frac{100}{N_{sp}}\times\frac{1\; sec}{\tau_{_{1/2}}}\right)^{2/3}\times$$
\begin{equation}
\label{fl1} 
\times\frac{R_{Astro}}{0.9 yr^{-1}}\times\frac{f_{Biotech}}{1}\times\frac{\mathcal{L}_{_{I}}}{t_{_{I}}}\;GeV\;s^{-1}\; cm^{-2},
\end{equation}
$$F^{\nu}_{_{II}}\simeq 2.3\times 10^{-8}\times\frac{100}{N_{sp}}\times$$
\begin{equation}
\label{fl2} 
\times\frac{R_{Astro}}{0.9 yr^{-1}}\times\frac{f_{Biotech}}{1}\times\frac{\mathcal{L}_{_{II}}}{t_{_{II}}}\;GeV\;s^{-1}\; cm^{-2}.
\end{equation}

Considering a Type-III civilization, for the flux from a nearest galaxy one obtains
$$F^{^{^{\nu}}}_{_{III}}\simeq 9.2\times 10^{-6}\times\frac{100}{N_{sp}}\times$$
\begin{equation}
\label{fl3} 
\times\left(\frac{R_{Astro}}{0.9 yr^{-1}}\times\frac{f_{Biotech}}{1}\times\frac{\mathcal{L}_{_{III}}}{t_{_{III}}}\right)^{2/3}\;GeV\;s^{-1}\; cm^{-2}.
\end{equation}
We have taken into account that an average distance between closest galaxies is given by $\left(V_g/N_g\right)^{1/3}$, where $V_U\simeq 1.22\times 10^{13}$ Mpc$^3$ is the volume of the visible universe and $N_g\simeq 10^{12}$ is an average number of galaxies in the universe.

As we have already explained, for Type-I societies the best way is to capture a single event, whereas for Type-II and Type-III civilizations upper limits of possible fluxes will be defined by a net effect of all quantum processing events every second. An interesting feature of Eqs. (\ref{fl2},\ref{fl3}) is that they do not depend on  $\tau_{_{1/2}}$, or black hole mass. Such a behaviour is a direct consequence of the fact that $n_x\sim\tau_{_{1/2}}/M\sim\tau_{_{1/2}}^{2/3}$, multiplied by $L^{\nu}_m\sim\tau_{_{1/2}}^{-2/3}$ (see Eq. \ref{power}) cancels the dependence on $\tau_{_{1/2}}$.

The IceCube sensitivity to detecting muon neutrinos is of the order of $F^{\nu}_{IC,min}\simeq 7\times 10^{-9}\;GeV\;s^{-1}\; cm^{-2}$ \citep{icecube}, therefore, the Type-I alien society cannot be detected for the mentioned parameters and it will be possible only for $f_{Biotech}\times \mathcal{L}_{_{I}}/t_I\simeq 7\times 10^4$, which seems to be an unrealistically large value. 

On the contrary, Type-II and Type-III alien societies might be visible to the Ice Cube detectors. Therefore, it is reasonable to estimate upper limits of the number of civilizations which might provide the flux for $>30$ TeV: $F^{\nu}_{IC}\simeq 3.6\times 10^{-7}\;GeV\;s^{-1}\; cm^{-2}$ observed by the IceCube instruments. By assuming that $F^{\nu}_{IC}$ is produced by the advanced ETI, one obtains

\begin{equation}
\label{n22} 
N^{\nu}_{_{II,max}}\simeq\frac{D_{MW}^2 F^{\nu}_{IC}}{n_{_{II}} L_m^{\nu}} \simeq 4.2\times 10^4,
\end{equation}
\begin{equation}
\label{n33} 
N^{\nu}_{_{III,max}}\simeq V_U\times\left(\frac{36 F^{\nu}_{IC}}{13 \pi n_{_{II}} L_m^{\nu}}\right)^{3/2}  \simeq 1.4\times 10^8,
\end{equation}
where we have taken into account that a flux from the nearest Type-III civilization should be multiplied by $13\pi^2/9$ to obtain the net effect from all civilizations uniformly distributed over the observed universe. Similarly, by taking $F^{\nu}_{IC,min}$,  into account, one can obtain the minimum values of civilizations, which might be detectable by the IceCube instruments: $N^{\nu}_{_{II,min}}\simeq 2.5\times 10^3$, $N^{\nu}_{_{III,min}}\simeq 2\times 10^6$.

Other signatures also might exist if ETI does not need to screen out the black hole radiation. In this case other particles will escape a region. In particular, if one considers escaping photons ($s = 1$ and $\alpha_1 = 20.4$, see Eq. (\ref{rate})), one can straightforwardly show that the spectral power peaks at

\begin{equation}
\label{en1} 
E^{\gamma}_m\simeq\frac{2.82 \hbar c^3}{8\pi GM}\simeq 43\times\left(\frac{100}{N_{sp}}\times\frac{1\; sec}{\tau_{_{1/2}}}\right)^{1/3} TeV,
\end{equation}
corresponding to the following gamma ray luminosity 
$$L^{\gamma}_m\simeq (E^{\gamma}_m)^2R(E^{\gamma}_m,M)\simeq$$
\begin{equation}
\label{power1} 
\simeq 1.5\times 10^{27}\times\left(\frac{100}{N_{sp}}\times\frac{1\; sec}{\tau_{_{1/2}}}\right)^{2/3}\; erg/s.
\end{equation}

By taking the number of all civilizations in the galaxy into account, one can derive the high energy photon fluxes for Type-I,II,III ETIs
$$F^{\gamma}_{_{I}}\simeq 1.2\times 10^{-14}\times\left(\frac{100}{N_{sp}}\times\frac{1\; sec}{\tau_{_{1/2}}}\right)^{2/3}\times$$
\begin{equation}
\label{fl11} 
\times\frac{R_{Astro}}{0.9 yr^{-1}}\times\frac{f_{Biotech}}{1}\times\frac{\mathcal{L}_{_{I}}}{t_{_{I}}}\;GeV\;s^{-1}\; cm^{-2},
\end{equation}
$$F^{\gamma}_{_{II}}\simeq 2.8\times 10^{-9}\times\frac{100}{N_{sp}}\times\frac{R_{Astro}}{0.9 yr^{-1}}\times$$
\begin{equation}
\label{fl22} 
\times\frac{f_{Biotech}}{1}\times\frac{\mathcal{L}_{_{II}}}{t_{_{II}}}\;GeV\;s^{-1}\; cm^{-2},
\end{equation}

$$F^{^{^{\gamma}}}_{_{III}}\simeq 1.1\times 10^{-6}\times\frac{100}{N_{sp}}\times$$
\begin{equation}
\label{fl33} 
\times\left(\frac{R_{Astro}}{0.9 yr^{-1}}\times\frac{f_{Biotech}}{1}\times\frac{\mathcal{L}_{_{III}}}{t_{_{III}}}\right)^{2/3}\;GeV\;s^{-1}\; cm^{-2}.
\end{equation}

The Major Atmospheric Imaging Cherenkov (MAGIC) telescope aiming to detect very high energy (VHE) gamma rays has sensitivity of the order of $10^{-13} erg\; cm^{-2}\; s^{-1}\simeq 6\times10^{-11} GeV\; cm^{-2}\; s^{-1}$   \citep{felix}. Therefore, MAGIC can detect the VHE gamma rays from Type-I civilizations if $\mathcal{L}_{_{I}}\sim 5\times 10^3 t_{_{I}}\sim 5\times 10^6$ yrs. 

From Eqs. (\ref{fl22},\ref{fl33}) it is evident that for the chosen parameters the VHE photons from Type-II  and Type-III alien societies might be detected by the MAGIC instruments. But one should clearly realise that if the advanced civilizations do not manufacture the black holes at the maximum rate (see Eqs. (\ref{n1}-\ref{n3})) the fluxes will be smaller. In any case, the sensitivity of MAGIC facilities impose a constraint on the minimum number of civilizations $N^{\gamma}_{_{II,min}}\simeq 60$, $N^{\gamma}_{_{III,min}}\simeq  7.5\times10^3$. 

\section{ETI from Dark Side} 

   All currently known elementary particles are described 
     by the Standard Model plus Einstein gravity. 
  The known particle species consist of:  the Higgs scalar,  
 the fermions (quarks and leptons) and the mediators of gravitational and gauge interactions. Counting all polarizations, this amounts 
 to $N_{\rm sp} \sim  100$ field theoretic particle species.   
  
  The Standard Model is an extraordinarily successful theory. 
  It suffices to say that within the range of distances of approximately 
  $10^{-17} - 10^{28}$cm  there is no experimental evidence 
  invalidating it.  Yet,  we know that this theory is incomplete and there must exist particle species beyond it.  
  
  Perhaps, the most 
  direct  evidence for this is provided by the dark matter.
  Although a part of dark matter could reside in black holes
  (for a review, see, \citep{BHDM}), 
  almost certainly,  this option too requires the existence of new particle species.  The current experimental constraints 
tell us that,  if lighter than $\sim$TeV, the new particle species 
cannot interact with us via the known forces other than gravity. 
A potential exception is provided by the hypothetical particles with  
very small electric charges.
 
Of course, new particles may interact with our species via some yet undiscovered forces. 
The only constraint on such interactions is 
that they are sufficiently weak for avoiding the 
experimental detection.  

 Nothing much beyond this is known about the hidden sector species. 
 They can come in number of interesting forms. For example, 
 dark sectors can be organized in form 
 of many hidden copies of the Standard Model \citep{DvaliN, DvaliRediN}, or the parallel branes separated from us in large 
 extra dimensions \citep{ADD}. 
 
 However, their number is limited.  
There exist an universal upper bound 
 \citep{DvaliN},
  \begin{equation}  \label{Nbound}
 N_{\rm sp} \sim 10^{32}  \,, 
\end{equation} 
on the total number of distinct particle species.  This bound comes from the fact that species lower the fundamental scale of quantum gravity to (\ref{Mbound}) 
\citep{DvaliN, DvaliRediN}.
Then, the lack of the observation of strong quantum gravity effects - such as 
the creation of micro black holes - at Large Hadron Collider, puts the lower bound $M_* \gtrsim $TeV.  This translates in the upper bound 
(\ref{Nbound}) on the total number of particle species.
This number of course includes the number of the known particle species of the Standard Model. 

     Currently, there are no model-independent restrictions on the ways the hidden  species organize themselves in different sectors and on how they interact with one another.   Given this vast freedom and our almost zero knowledge of 
  the possible forms and the evolutionary paths    
 of intelligence, there is no reason 
   to restrict ETI exclusively to the creatures consisting of 
   particles of the Standard Model.   
     Thus, with our current knowledge there is a room for the existence of $\sim 10^{32}$ distinct dark  sectors each producing more than one type of intelligent beings.  
   
On this extended landscape of possibilities, the black holes 
offer to provide an universal search tool. First, as explained, 
due to their maximal efficiency, the black hole based
devices represent universal attractor points in the evolution of     
quantum computing. 
  Secondly, the usage of black holes as the information storing devices can produce the observable signatures of the dark ETI. Due to the fact that Hawking radiation is democratic in particle species, the dark 
 quantum computers will inevitably radiate the species of the Standard Model, such as neutrinos and photons. 
 
  Some comments are however in order.  
 If the number of species $N_{\rm sp}$ is significant,
 the characteristics of the lightest black holes will be affected. 
 In pure Einstein gravity,  the quantum gravity becomes strong 
 at the Planck scale $M_P$.  That is, the interactions of particles 
 at distances $\sim L_P$ are fully dominated by gravity. 
 
 Correspondingly, 
 the Planck scale sets both the mass, $M \sim M_P$, as well as  
the radius, $R \sim L_P$, of the lightest black hole.  
No lighter black hole can exist in nature within Einstein 
theory of gravity.

This fact creates a technical obstacle for the civilizations seeking  to manufacture the light black holes. For example, in order to produce a $M_P$-mass 
black hole,  we need to collide some elementary particles with a center of mass energy 
$\sim M_P$ and an impact parameter $\sim L_P$. 
This is highly non-trivial.  
 For example, with the current technology available to our civilization, boosting particles to such high energies would require a particle accelerator of the cosmic extend. 

    However, in a theory with large number of particle species,
   the story is dramatically different.  The scale of strong gravity 
    is lowered to the scale $M_*$ given by (\ref{Mbound}). Correspondingly, 
    the mass of the lightest black hole is now given by 
    $M_*$, as opposed to $M_P$. 
  Such a low scale can be much easier accessible. 
    It suffices to notice that, for example, in the extreme case of $N_{\rm sp} \sim 10^{32}$, such a 
  black hole can even be manufactured by our current civilization at LHC \citep{DvaliRediN}. 
  
 In fact, a particular case of  this is provided  by a well known prediction \citep{AADD} of the theory of large extra dimensions introduced in \citep{ADD}.  
 In this theory, the role of additional particle species  is played 
 by the Kaluza-Klein tower of gravitons. Their number is precisely
 $N_{\rm sp} \sim 10^{32}$ \citep{DvaliN}.

   The general message is that with large $N_{\rm sp}$, the black holes smaller than a certain critical radius $R_*$ are lighter than their counterparts in pure Einstein gravity \citep{DvaliBHSP}.
 Below this size, the precise 
  relation between $M$ and $R$ is theory-dependent. 
  For example, in theory \cite{ADD} with $d$ extra dimensions of size  
  $R_*$, the black holes of radius $R < R_*$ and mass $M$ obey
  the following relation, 
    \begin{equation} \label{MR} 
   R^{1+d} \sim \frac{M}{M_*^{2+d}}  \,. 
\end{equation} 
  The theory-independent phenomenological bound on the critical 
  scale $R_*$ comes from the short distance tests of Newtonian gravity.  The current bound of $R_* \lesssim 38\mu$m
is provided by the torsion-balance experiments \citep{torsion}. 
   
  In the light of the above, we must take into account that for large number of species, 
   it becomes easier to manufacture the small black holes. 
   Correspondingly, with larger $N_{\rm sp}$, the compact storage and a fast processing of  information via black holes becomes more accessible even for less advanced ETIs.

\section{Summary}

 The advancement of civilization is directly linked with its ability to efficiently 
 store and process information. 
    It is therefore important to understand what are the fundamental 
 limitations of this advancement imposed by our 
 current understanding of laws of nature.  At most fundamental 
 and experimentally best-verified level, 
 this understanding 
 comes from the framework of quantum field theory.  
  
     Recent  studies \citep{DvaliSaturon} show that
  the validity of quantum field theory imposes an universal
upper bound on the information storage capacity
of a device that can be composed by the corresponding quanta.  The objects saturating this bound, 
the so-called "saturons",  are the most efficient storers of quantum information.  

Interestingly, the universality of the gravitational interaction
tells us  that, among all possible hypothetical saturons, the
black holes have the highest capacity of the information storage \citep{DvaliSaturon}.  

  Correspondingly, any sufficiently advanced civilization,  
 is ultimately expected to develop the black hole based
 quantum computers.  This opens up an exciting prospect 
 for SETI through the Hawking radiation from such computers. 
 
  Certain generic features emerge. 
  
    The estimated efficiency of information processing indicates that 
    "memory disks"  that allow for short time (e.g., $t_q \lesssim$ sec)
     of information retrieval 
   must be based on microscopic black holes (e.g., of size, $R \lesssim 10^{-18}$cm).    Correspondingly,  the Hawking radiation from ETI 
 must be in a high energy spectrum that is potentially detectable by the human devices.

We would like to make emphases on the following point.  As we have argued, it is reasonable to assume that the computation time for advanced ETI is of order or shorter than a second. The black holes that are able to provide such a short information retrieval time are lighter than $10^9$ g. This provides an unique smoking gun for our idea of SETI, since the signals coming from the evaporation of such artificial black holes have no competitors among the natural ones. The reason is that naturally-produced black holes of the masses $<10^{15}$g must be exclusively primordial, since no available natural mechanisms exist for their production in contemporary  Universe. The primordial black holes with such a low mass would have evaporated long ago. This leaves us with an exciting option that any Hawking signal in the proposed energy range must come from a black hole that is artificially manufactured.

    Even if ETI is entirely composed of  totally new 
    types of particle species, due to universality of Hawking radiation, the emitted quanta  
    will contain the known particles, such as neutrinos or photons.  
    Thus, the ultimate efficiency of black hole quantum computing 
    allows us to search for ETI even if they are entirely composed of 
    some unknown 
    quanta. 
    
      The main message of our paper is that a new knowledge 
     about ETI can be deduced by analysing    
      the  Hawking radiation  from their quantum computers. 
      In turn, the non-observation of such a radiation, can be translated 
      as new limits on ETI.  
      
    We gave some indicative estimates.  Analysing the IceCube's VHE ($>30$ TeV) data we showed that the mentioned 
    instruments are able to detect the neutrino emission flux from black hole  farms of Type-II,III 
    advanced civilizations. In particular, the current sensitivity and the observed VHE flux, 
    impose constraints on the numbers of civilizations $2.5\times 10^3\leq N^{\nu}_{_{II}}\leq 4.2\times 10^4$, $2\times 10^6\leq N^{\nu}_{_{III}}\leq 1.4\times 10^8$.
             
  We have also pointed out that need for the black hole 
  computers can lead to accompanying signatures from the
  manufacturing process. For example, black hole factories 
  will result into a radiation from the high energy collisions that 
  are necessary for formation of micro black holes.
  
    Although, most likely, our knowledge covers only a limited 
  number of the existing particle species, the above features 
  are independent on this lack of knowledge.     
  Morever, a detection of Hawking radiation from ETI, can 
  indirectly contribute to widening of our knowledge about hidden particles. 
   This is due to a general fact that larger is the number 
   of particle species $N_{\rm sp}$, easier it becomes 
   for any  civilization to manufacture black holes
   \citep{DvaliRediN}. 
   This is true also for our civilization, which has a chance of 
creating micro black holes at colliders, provided $N_{\rm sp} \sim 10^{32}$. 

  In summary, the universality of the laws of quantum physics and gravity 
  allows us to draw some surprisingly powerful conclusions 
  about ETI.  
   We have argued that on the evolution road of any 
 long-lasting civilization  the black hole based 
   quantum computers are the natural attractor points. 
    This opens up a new avenue for SETI.
 
  The general smoking guns include potential detections
  of high energy neutrino fluxes with approximately thermal 
 distribution.
  Another interesting but more exotic possibility, that requires additional clarifications, is that farms of 
 closely spaced black holes which likely produce and extended 
 high temperature medium may occasionally 
 create and radiate away heavy composite structures.

\section*{Acknowledgments}

The work of G.D. was supported in part by the Humboldt Foundation under Humboldt Professorship Award, by the Deutsche Forschungsgemeinschaft (DFG, German Research Foundation) under Germany's Excellence Strategy - EXC-2111 - 390814868, and Germany's Excellence Strategy under Excellence Cluster Origins.
  Z.O. would like to thank  Arnold Sommerfeld Center, 
	Ludwig Maximilians University and Max Planck Institute for Physics
	for hospitality during the completion of this project. 

\section*{Data Availability}

Data are available in the article and can be accessed via a DOI link.

\end{document}